\documentclass{article}
\usepackage{a4}
\usepackage{graphicx}
\begin{document}

\title{GUT angle minimises $Z^0$ decay}
\author{Alejandro Rivero\thanks{Zaragoza University at Teruel.  
           {\tt arivero@unizar.es}}}

\maketitle

\begin{abstract}
The GUT value of Weinberg's angle is also the value that minimises the total
square matrix elements of $Z^0$ decay,
independently of any GUT consideration, and thus the one that maximises
the neutrino branching ratio against total width. We review the proof of
this result and some related facts.
\end{abstract}

From any textbook (eg \cite{Ellis:1991qj}), the amplitude for decay of $Z^0$ into
a fermion pair, at leading order, is
\begin{equation}
\Gamma(Z^0\to f\bar f)=C_f (|V_f|^2+|A_f|^2) {G_F M_Z^3 \over 6 \sqrt 2 \pi}
\end{equation}
where $C_f$ is a colour normalisation constant, 1 for fermions and 3 for quarks, and
$V_f$ and $A_f$ are the vector and axial charges,
\begin{eqnarray}
V_f&=&T^3_f -2 Q_f \sin^2 \theta_W \\
A_f&=&T^3_f
\end{eqnarray}

If we are interested on the decay into a set of fermions, we add the
contributions to get the total square matrix element:
\begin{equation}
K_{\{f\}}=\sum_f C_f ((T^3_f)^2+(T^3_f -2 Q_f \hat s)^2)
\end{equation}

We want to know for which value of $\hat s \equiv \sin^2 \theta_W$ will the 
relative coupling, and then the decay width\footnote{Except for the kinematics if
we still want $M_Z$ to depend of $\sin\theta_W$.}, to be a minimum. Thus we ask
\begin{equation}
0=K_{\{f\}}'(\hat s)= \sum_f 2 C_f (T^3_f -2 Q_f \hat s) ( -2 Q_f)
=4 \sum_f C_ f (2 \hat s Q_f^2 -T^3_f Q_f)
\end{equation}
and then using that $T^3_f Q_f = T^3_f (Y+T^3_f)= (T^3_f)^2$ for sums across
an isospin multiplet, accounting colour
in the whole sum, and passing it from Dirac to Weyl species we get
\begin{equation}
\hat s_{min}= {\sum_f T^3_f Q_f \over 2 \sum_f Q_f^2} =
 {\sum (T^3_f)^2 \over \sum Q_f^2}
\end{equation}

When the set of fermions is a whole generation, this last formula equals the very
well known result (e.g. exercise VII.5.2 in \cite{Zee:2003mt}) for
$\sin^2 \theta_W$ at the GUT scale of any unification based
on a simple group. It is independent of the specific fermion content of the theory
except that the factor 2 cancels because  right fermions live in isospin
singlets, thus $T^3=0$ for them.

The proof is even shorter if we make use of the argument from  D. H. 
Perkins \cite[section 9.3]{Perkins:1982xb}. According this argument, 
normalisation of the coupling amplitudes in the GUT scale amounts
to forbid fermion-loop mixing of $Z^0$ 
and photon, i.e. to ask directly $\sum_{f} C_f Q (T^3 - 2 \hat s \cdot Q ) =0$ when
the loop is run for all the fermions in a multiplet of the GUT group. Diagramatically
we can move between our formulation and Perkins's by joining the two fermion lines
in the $Z^0$ decay to form a tadpole diagram and then inserting an external
photon line, the whole process resembling the setup of a Ward identity, but where now
the derivative is against Weinberg's sine squared. 

Another early try to get the right normalisation of 
couplings from considerations in the fermion-boson vertex can be found in \cite{Rosen:1974ye}.

Now lets concentrate particularly in the fermion content of a generation
of the standard model, where
we have $\hat s_{min}=3/8$ (and $K_{\{u,d,\nu,e\}}(3/8)=2.5$). 

The charge assignments of the standard model can be imposed by hand or via 
the requisites of anomaly cancellation. In any case, they have an extra property
when we pay attention to minimisation of $Z^0$ decay:  the value $3/8$ also
minimises separately the partial decay width towards an $u$ (or $c$) quark. And thus
it minimises also the partial decay into the set $\{d,\nu_e,e\}$ (or $\{s,\nu_\mu,\mu\}$
or $\{b,\nu_\tau,\tau\}$)

This means that for the standard model, 3/8 is not only the value minimising decay
into first and second generation; it also minimises the decay of $Z^0$ into
the third generation, even if the top quark is kinematically out of reach of
the gauge meson. A posteriori, we could interpret this fact of an indication of the
particular characteristics of the top quark. The same arguments could be run from
Perkins requisite, but the sum of fermion loops does not underline the special
role of top quark, while $Z^0$ decay stresses it. 

Up to here the main comment, or result\footnote{The whole note is motivated because I have
been unable to find this remark in standard textbooks; I'd thank any information about
previous statements of it}, of this note: that the GUT formula for Weinberg angle at
unification scale is also got without GUT, by minimising $Z_0$ decay. The following few paragraphs
are random musings distilled from the above:

- A consequence of the derivation here presented is that a model can
get into GUT angle by asking for some minimisation requisite, without looking
for a GUT group. Spectral actions of Connes-Chamseddine could be a good candidate
for this, as would some other approaches from non commutative geometry \cite{CamCoq,Scheck:1992pt}.
And I wonder if Iba\~nez string-inspired approaches to Weinberg angle
are also a consequence of hidden minimisation.

- We have an alternate, more physical if you wish, way to state the Hierarchy problem. 
Instead of asking "Why the unification scale is so high compared with the electroweak
SSB scale", now we can ask "Why the Z0 decay (the squared matrix elements $K$) should
reach a minimum at the GUT scale". 

- If we contemplate $K_{\{u,d,\nu,e\}}$ we can wonder for the value of this coupling
at the experimental scale of decay, ie when $\hat s$ is about 0.232. In GUT theories
there is an scale available from which the value of the angle descends via renormalisation
flow \cite{georgi,Dimopoulos:1981yj}; here we haven't such scale to start with.
But a related unexplained fact is that
\begin{equation}K_{\{u,d,\nu,e\}}(0.231948...)=\exp(1)=\sum_0^\infty {1 \over n!}\end{equation}
We haven't the slightest idea of why the transcendent number e could have a reason to
appear here. The minimum, $K=2.5$, is a member of the simple series expansion of e, up to
three terms. But on the other hand the values $K=1$ and $K=2$, which we could get by using
the lower terms, need of a complex $\hat s$.
 
It could bring some problem to poorly programmed statistical algorithms.
Lets trace numerically this dependence $K_{\{f\}}(\hat s^2)$
$$
\begin{array}{ll}
\hat s^2 &   | \ln K_{\{f\}} -1 |  \\
0.2310 &  .001067 \\
0.2311 &  .000954 \\
0.2312 &  .000841 \\
0.2313 &  .000729 \\
0.2314 &  .000616 \\
0.2315 &  .000503 \\
0.2316 &  .000391 \\
0.2317 &  .000279 \\
0.2318 &  .000166 \\
0.2319 &  .000054 \\
0.2320 &  .000059 \\
0.2321 &  .000171 \\
0.2322 &  .000283 \\
0.2323 &  .000395 \\
0.2324 &  .000507 \\
0.2325 &  .000619 \\
0.2326 &  .000731 \\
0.2327 &  .000842 \\
0.2328 &  .000954 \\
0.2329 &  .001065
\end{array}
$$

An algorithm using a three-digits cutoff somewhere (say, in a conditional IF
of the simulation code) could round the values between 0.2318 .. 0.2320 
into the value 0.2319484

In any case, compare with $0.23193 \pm 0.00056$ from ALEPH \cite{aleph} hep-ex/0107033. It is amusing.

{\centering
 \includegraphics[width=12.8cm]{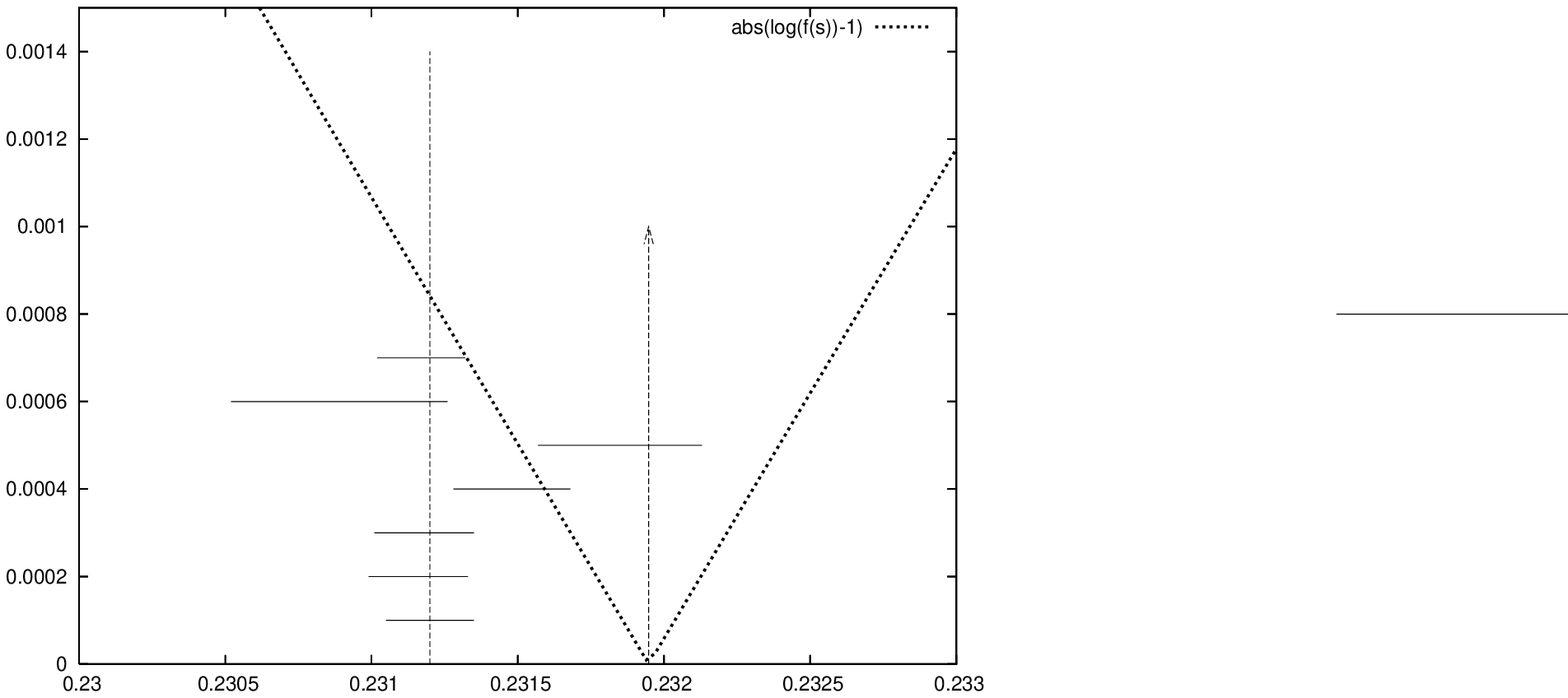}
 
 {The figure shows $|1-\ln K|$ as in the table, compared with the measured
 values of $\hat s^2$ from table 10.5 of \cite{pdg}. Vertical arrows mark experimental
 central value, for all data, and the point 0.2319484 of singularity. It can be seen that
 the measurement from $A^{b,c}_{FB}$ is away from the rest of values but in agreement with
 the numerical singularity. }}

But the precise apparition of the trascendent number $e$ or its series should be taken with a bit 
of salt if thinking about applications in phenomenology. To give one example, almost the same 
numbers (0.2319478) are  got if we "solve"
$e$ from its approximation $$\sqrt{e-5/2} \approx (1+3/8) {e \over 8} $$
I.e., if we ask the derivative of $\ln K$ to
have the value $$\left.{d(\ln K(\hat s^2)) \over d(\hat s^2)}\right|_{\hat s^2_Z} \approx -\sqrt{\frac 23}(1+\frac 38)$$
which give us a intrascendent (er, algebraic) value. And for sure other approximations are possible.

Related to this, it is perhaps worth to mention that the experimental $Z^0$ width, which 
we can calculate by summing the three families, has another intriguing empirical accident: 
it scales straightforwardly, via the cube of the mass, from neutral pion
width\cite{Rivero:2005ky}, so that currently the scaling down from $Z^0$ to $\pi^0$ is a precise
"prediction" of the mean life of the latter particle. 

- Also for the standard model assignment of charges, and in terms of the decay to a 
whole family, we have the relationships $K_{\{d,e\}}=\frac 12 K$, $K_{\nu}=\frac 12$,
$K_{u}=\frac 12 K - \frac 12$. This implies that for the above mentioned $K=1$, the
decay probability into upper quarks vanishes. 

- The value of $K_{\nu}$ does not depend of $\sin^2 \theta_W$. Then we can use its
corresponding decay width to stablish branching ratios, getting rid of $G_F$ and 
$M_{Z^0}^3$. So, we can state for instance that at tree order, the value of $\sin^2 \theta_W$
that maximises the branching ratio of neutrinos in $Z^0$ decay is the same value
that $\sin^2 \theta_W$ has at the Grand Unification Scale. And so on. This is 
a traditional trick, used by instance in \cite{Wilczek:1975kn}, where regretly
Wilczek et al. only plot ratios dependence of $\sin^2 \theta_W$ in the 
quark sector, then getting a near miss of the formulation here presented.

- Just for analytic commodity 
we can solve the equation for $\hat s$ in terms of the
decay to a whole family. We have
\begin{equation}
\hat s= \frac 38 (1- \sqrt \frac 23 \sqrt{K-\frac 52})
\end{equation}
where some care must be taken about the lack of analyticity of $||^2$ and
its conversion to $()^2$. Actually the possible values of $\hat s$ for a given K
form an hyperbola in the complex plane, that becomes degenerate when $K=5/2$. For
values $K<5/2$ the hyperbola does not touch the real axis and the above equation
gives the position of its vertex. Incidentally, for K=1 such vertex it at 
a distance 3/8 of the real axis and the hyperbola is symmetric to the one 
for K=4.

- Finally, let me note that another common apparition of the factor 3/8 is in
perturbative expansions of electromagnetism, and that the use of the GUT
to correct $\alpha_{EM}$ has been vindicated in some exponential adjustments 
between Planck and electron scales, eg by Laurent Nottale. I strongly doubt
that the development here can be connected to these ones, albeit a corner 
should be left to accidental technicalities from $\sin^4, \cos^4$ expansions.


\begin{thebibliography}{30}

\bibitem{Ellis:1991qj}
  R.~K.~Ellis, W.~J.~Stirling and B.~R.~Webber,
  { ``QCD and collider physics,''}
  Camb.\ Monogr.\ Part.\ Phys.\ Nucl.\ Phys.\ Cosmol.\  {\bf 8}, 1 (1996).
%%CITATION = CMPCE,8,1;%%

\bibitem{Zee:2003mt}
  A.~Zee,
  ``Quantum field theory in a nutshell,''
%\href{http://www.slac.stanford.edu/spires/find/hep/www?irn=5722500}{SPIRES entry}
  Princeton University Press (2003).

\bibitem{georgi}
  H.~Georgi, H.~R.~Quinn and S.~Weinberg,
  ``Hierarchy Of Interactions In Unified Gauge Theories,''
  Phys.\ Rev.\ Lett.\  {\bf 33} (1974) 451.
  %%CITATION = PRLTA,33,451;%%


%\cite{Dimopoulos:1981yj}
\bibitem{Dimopoulos:1981yj}
  S.~Dimopoulos, S.~Raby and F.~Wilczek,
  ``Supersymmetry And The Scale Of Unification,''
  Phys.\ Rev.\ D {\bf 24} (1981) 1681.
  %%CITATION = PHRVA,D24,1681;%%
  
  
%\cite{Ma:1977da}
\bibitem{Ma:1977da}
  E.~Ma,
  %``Eigenvalue Condition For The Weinberg Angle And Possible New Leptons And
  %Quarks,''
  Prog.\ Theor.\ Phys.\  {\bf 58}, 1896 (1977).
  %%CITATION = PTPKA,58,1896;%%
    
%\cite{Perkins:1982xb}
\bibitem{Perkins:1982xb}
  D.~H.~Perkins,
  ``Introduction To High-Energy Physics,'' 4th edition
%\href{http://www.slac.stanford.edu/spires/find/hep/www?irn=1064908}{SPIRES entry}
  
%\cite{Rosen:1974ye}
\bibitem{Rosen:1974ye}
  S.~P.~Rosen,
  %``Weinberg Angle As A Test Of The Georgi-Glashow Unified Gauge,''
  Phys.\ Rev.\ Lett.\  {\bf 33} (1974) 614.
  %%CITATION = PRLTA,33,614;%%

%\cite{Rivero:2005ky}
\bibitem{Rivero:2005ky}
  A.~Rivero,
  ``Anomaly-driven decay of massive vector bosons,''
  arXiv:hep-ph/0507144.
  %%CITATION = HEP-PH 0507144;%%

\bibitem{CamCoq}
  G.~Cammarata and R.~Coquereaux,
  ``Comments about Higgs fields, noncommutative geometry and the standard
  model,''
  arXiv:hep-th/9505192.
  %%CITATION = HEP-TH 9505192;%%
  
  %\cite{Scheck:1992pt}
\bibitem{Scheck:1992pt}
  F.~Scheck,
  ``Anomalies, Weinberg angle and a noncommutative geometric description of the
  Standard Model,''
  Phys.\ Lett.\ B {\bf 284} (1992) 303.
  %%CITATION = PHLTA,B284,303;%%
  
\bibitem{Wilczek:1975kn}
  F.~A.~Wilczek, A.~Zee, R.~L.~Kingsley and S.~B.~Treiman,
  ``Weak Interaction Models With New Quarks And Righthanded Currents,''
  Phys.\ Rev.\ D {\bf 12} (1975) 2768.
  %%CITATION = PHRVA,D12,2768;%%
  

\bibitem{pdg}
S. Eidelman et al., Physics Letters B592, 1 (2004) 
 and 2005 partial update for edition 2006 

\bibitem{aleph}
  A.~Heister {\it et al.}  [ALEPH Collaboration],
  ``Measurement of A(FB)(b) using inclusive b-hadron decays,''
  Eur.\ Phys.\ J.\ C {\bf 22} (2001) 201
  [arXiv:hep-ex/0107033].
  %%CITATION = HEP-EX 0107033;%%


\end{thebibliography}
\end{document}